\documentclass[conference, 10pt,a4paper,twocolumn]{IEEEtran}
\IEEEoverridecommandlockouts
\usepackage{amsfonts}
\usepackage{colortbl}
\usepackage[a4paper, total={184mm,239mm}]{geometry}
\usepackage{subcaption} 
\usepackage{threeparttable}
\usepackage{arydshln}
\usepackage{siunitx}
\usepackage{pgfplots} 
\usepackage{fancyhdr}
\fancypagestyle{IEEE}{\fancyhf{}\vspace{-0.8cm}\fancyfoot[C]{\small © 2024 IEEE. Personal use of this material is permitted. Permission from IEEE must be obtained for all other uses, in any current or future media, including reprinting/republishing this material for advertising or promotional purposes, creating new collective works, for resale or redistribution to servers or lists, or reuse of any copyrighted component of this work in other works.\\ This paper was published in \textit{M. Brunner, H. H. Lee, A. Hepp, J. Baehr and G. Sigl, "Hardware Honeypot: Setting Sequential Reverse Engineering on a Wrong Track," 2024 27th International Symposium on Design \& Diagnostics of Electronic Circuits \& Systems (DDECS), Kielce, Poland, 2024, pp. 47-52, doi: 10.1109/DDECS60919.2024.10508924. https://ieeexplore.ieee.org/document/10508924}}}
   
\usepackage{acro}
\acsetup{trailing/activate}
\DeclareAcroEnding{subst}{}{er}
\DeclareAcronym{RTL}{short = RTL, long = Register Transfer Level}
\DeclareAcronym{RE}{short = RE, long = Reverse Engineering}
\DeclareAcronym{PCA}{short = PCA, long = Principal Component Analysis}
\DeclareAcronym{FSM}{short = FSM, long = Finite State Machine}
\DeclareAcronym{HT}{short = HT, long = Hardware Trojan}
\DeclareAcronym{GNN}{short = GNN, long = graph neural network}
\DeclareAcronym{FF}{short = FF, long = Flip-Flop}
\DeclareAcronym{SFF}{short = state FF, long = State Flip-Flop}
\DeclareAcronym{STG}{short = STG, long = State Transition Graph}
\DeclareAcronym{FSM-HP}{short = FSM-HP, long = Finite State Machine Honeypot}
\DeclareAcronym{FP}{short = FP, long = Feedback Path}
\DeclareAcronym{SCC}{short = SCC, long = Strongly Connected Component}

\usepackage[style=ieee,backend=biber,isbn=false,doi=false,eprint=false,maxcitenames=2,mincitenames=1,minbibnames=1,maxbibnames=6]{biblatex}
\DeclareFieldFormat{url}{\url{#1}}
\DeclareFieldFormat{sentencecase}{#1}
\usepackage{xpatch}
\xpatchbibdriver{online}
{\printtext[parens]{\usebibmacro{date}}}
{\iffieldundef{year}{}{\printtext[parens]{\usebibmacro{date}}}}

\begin{document}

\title{Hardware Honeypot: Setting Sequential Reverse Engineering on a Wrong Track
\thanks{This work was partly sponsored by the Federal Ministry of Education and Research of Germany in the project VE-FIDES under Grant No.: 16ME0257}
}

\author{\IEEEauthorblockN{Michaela Brunner\IEEEauthorrefmark{1}, Hye Hyun Lee\IEEEauthorrefmark{1}, Alexander Hepp\IEEEauthorrefmark{1}, Johanna Baehr\IEEEauthorrefmark{2}, Georg Sigl\IEEEauthorrefmark{1}\IEEEauthorrefmark{2}}
\IEEEauthorblockA{\IEEEauthorrefmark{1}\textit{TUM School of Computation, Information and Technology, Chair of Security in Information Technology}\\
\textit{Technical University of Munich}, Munich, Germany \\
michaela.brunner@tum.de, hyehyun.lee@tum.de, alex.hepp@tum.de, johanna.baehr@aisec.fraunhofer.de, sigl@tum.de}
\IEEEauthorblockA{\IEEEauthorrefmark{2}\textit{Fraunhofer Institute for Applied and Integrated Security (AISEC)}, Garching b. Munich, Germany}
}

\maketitle

\begin{abstract}
Reverse engineering (RE) of finite state machines (FSMs) is a serious threat when protecting designs against RE attacks. While most recent protection techniques rely on the security of a secret key, this work presents a new approach: hardware FSM honeypots. These honeypots lead the RE tools to a wrong but, for the tools, very attractive FSM, while making the original FSM less attractive. The results show that state-of-the-art RE methods favor the highly attractive honeypot as FSM candidate or do no longer detect the correct, original FSM.
\end{abstract}


\section{Introduction}
\label{sec:intro}
\thispagestyle{IEEE}

\ac{RE} is a severe threat in the silicon supply chain, endangering the intellectual property's reliability, confidentiality, and integrity. In particular, the \ac{FSM} of the design is of special interest to an attacker because it reveals
what constitutes the design: the design's functionality.
To prevent \ac{RE} of \acp{FSM}, \ac{FSM} obfuscation methods which lock the functionality with secret keys, with dynamically changing keys, or based on input patterns may be used, e.g. \cite{janusHD,rahman2022practical,keyless2023}.
Obfuscation schemes without a potentially attackable locking key are an alternative, like camouflaging-based methods.
Camouflaging, however, often requires a foundry with special capabilities to implement it into the design, like adding a thin isolating layer to gate contacts \cite{hoffmann2021doppelganger}; or the designer has to reveal the camouflaged information, like the gate and wire delays \cite{brunner2022TimingCamouflage}, to the foundry for manufacturing. 
This work, in contrast, presents a new technique that is not based on foundry-enabled camouflaging or locking.
It hinders state-of-the-art \ac{RE} methods from successfully identifying the entire set of correct \ac{FSM} gates in a gate-level netlist by purposefully exploiting characteristics of \ac{RE} methods, which results in a lack of necessary information, similar to camouflaging.
In addition and similar to \cite{hoffmann2021doppelganger}, it leads the attacker to a wrong, designer-controlled \ac{FSM}.

To extract an \ac{FSM} in a gate-level netlist, sequential \ac{RE} methods exist. These methods first identify the \acp{FF} of the \ac{FSM}, so-called \acp{SFF}, and all other combinational gates belonging to the \ac{FSM}.
Next, they extract the \ac{STG} \cite{mcelvain2001methods}.
Many state-of-the-art sequential \ac{RE} methods do not fully investigate the extraction of multiple \acp{FSM} \cite{9794151}.
Some extract multiple \ac{FSM} candidates but do not further elaborate on how to choose the correct candidate \cite{shi2010highly, 3513086}.
Others extract only one \ac{FSM} candidate \cite{brunner2019improving, 9643498}.
Thus, the existence of multiple \acp{FSM} within a design complicates sequential \ac{RE}.
In addition, current \ac{SFF} identification methods are heuristic approaches that use specific---often similar---\underline{features} to identify \acp{SFF}.

\begin{table*}[t]
    \caption{Overview of \ac{SFF} features which are exploited by state-of-the-art \ac{SFF} identification methods.}
    \label{tab:BackgroundOverview}
    \centering
    \begin{tabular}{c|ccccccc}
         Method & \cellcolor{gray!10} High FP & Any FP & Grouping based on &  Effect on & \cellcolor{gray!10} Dissimilarity & Influence/dependency & Other structural \\
          & \cellcolor{gray!10}& & `clock' or `enable' or `reset' &  control signals & \cellcolor{gray!10}& behavior or SCC & features\\
        \hline
        \cite{mcelvain2001methods} & \cellcolor{gray!10}\checkmark &&&&\cellcolor{gray!10}&&\\
        \cite{shi2010highly} & \cellcolor{gray!10}\checkmark & & \checkmark & \checkmark & \cellcolor{gray!10}& \checkmark &\\
        \hdashline
        \cite{meade2016gate, meade2018old, 9218616} & \cellcolor{gray!10}& & & & \cellcolor{gray!10} \checkmark &  &\\
        \cite{meade2019neta, netadoku} & \cellcolor{gray!10}& & & & \cellcolor{gray!10} \checkmark & \checkmark & \checkmark\\
        \cite{brunner2019improving} & \cellcolor{gray!10}& \checkmark & & & \cellcolor{gray!10} \checkmark &&\\
        \cite{3513086} & \cellcolor{gray!10}& \checkmark & & & \cellcolor{gray!10} \checkmark & \checkmark & \\
        \hdashline
        \cite{fyrbiak2018difficulty} & \cellcolor{gray!10} \checkmark & & \checkmark & \checkmark  & \cellcolor{gray!10}& \checkmark & \\
        \cite{9643498} & \cellcolor{gray!10} & & &  & \cellcolor{gray!10}& \checkmark & \checkmark \\
    \end{tabular}
\end{table*}
\begin{figure}
    \begin{center}
		\includegraphics[width = 0.43\textwidth]{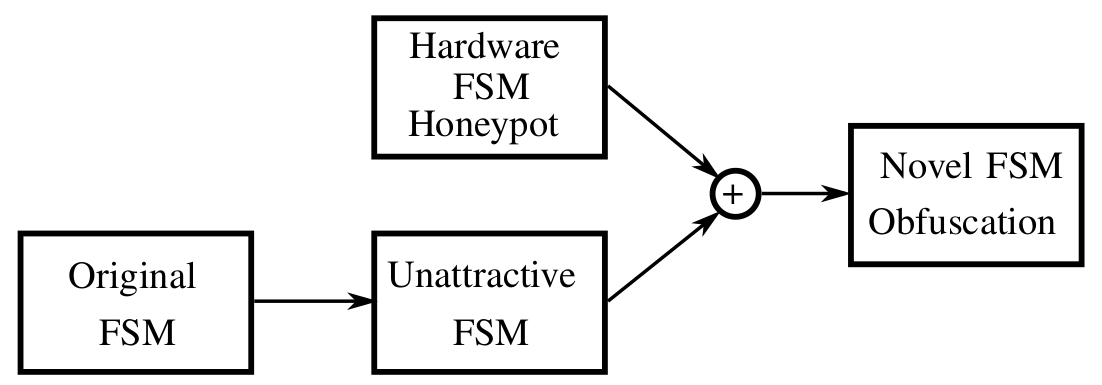}
	\end{center}
    \caption{Novel two-part \ac{FSM} obfuscation: hiding the original \ac{FSM} by making it less attractive (unattractive \ac{FSM}) and providing an attractive alternative (hardware \ac{FSM} honeypot)}
    \label{fig:HPMotivation}
\end{figure}
Using these two challenges, we present a novel two-part \ac{FSM} obfuscation methodology to prevent sequential \ac{RE}, see Fig. \ref{fig:HPMotivation}.
\begin{itemize}
    \item We introduce \textit{hardware \acp{FSM-HP}} which \underline{satisfy features} of the \ac{SFF} identification methods.
\acp{FSM-HP} pretend to be the correct \ac{FSM} of the design, which causes attackers to stop their effort to extract further \acp{FSM} and thus prevents the extraction of the correct \ac{FSM}.
To ensure that sequential \ac{RE} methods identify the \ac{FSM-HP} as a single or best \ac{FSM} candidate, we design it to be more attractive than the correct \ac{FSM}.
    \item We obfuscate the original \acp{FSM}, now called \textit{unattractive \acp{FSM}}, by \underline{eliminating features} of certain \ac{SFF} identification methods. As a result, unattractive \acp{FSM} are resistant to these identification methods. Similar approaches exist for example in the context of \acp{HT} \cite{9859423} or logic locking \cite{simLLDutta2023} which insert \acp{HT} or locking gates with weaker features of their detection or breaking methods to circumvent these attacks.
    \item We combine both techniques, \acp{FSM-HP} and unattractive \acp{FSM}, to enhance the effect of both.
\end{itemize}

\acp{FSM-HP} can be implemented on \ac{RTL} or gate level, allowing designers to control design properties or functionality. Thus, they can increase the attractiveness of the \ac{FSM-HP} or engage an attacker with controlled, false information.
The obfuscation results show that state-of-the-art \ac{SFF} identification methods either favor the \acp{SFF} of the \acp{FSM-HP} or can no longer identify the correct \acp{SFF}.
Both lead to a wrong \ac{FSM} extraction.
Our threat model assumes access to a reverse-engineered gate-level netlist and sequential \ac{RE} tools, but no access to further tools or design specifications, like simulation stimuli.
Thus, since attackers do not know the correct \ac{FSM}, they cannot expose an extracted, plausible-looking \ac{FSM-HP} as fake \ac{FSM}.

In the following, section \ref{sec:background} presents preliminaries and a background analysis. Section \ref{sec:methodology} presents the novel obfuscation, in particular \acp{FSM-HP} and unattractive \acp{FSM}. The results are discussed in section \ref{sec:Results}. Section \ref{sec:Conclusion} concludes the work.

\section{Background Analysis}
\label{sec:background}
To identify similar features that can be exploited to artificially modify the attractiveness of \acp{FSM}, we introduce preliminaries and summarize and compare \ac{FSM} extraction algorithms, in particular \ac{SFF} identification methods.

\subsection{Preliminaries}
The section introduces \acp{FSM} and further defines their structural characteristics, namely path properties and connectivity.
\subsubsection{FSM}
The \ac{STG} of an \ac{FSM} consists of a set of states, inputs, transitions, and a reset state \cite{moore1956gedanken}.
Synthesis translates the \ac{FSM} into a gate-level netlist.
State FFs or the state register hold the \ac{FSM} state, while their combinational input cone, the next state logic, updates the \ac{FSM} state for each clock cycle, implementing the state transitions.

\subsubsection{Path Properties} A connection from an \ac{FF} to itself is called a \ac{FP}. There exist three path strengths between \acp{FF} \cite{3513086}: A high strength path contains only combinational gates, a medium strength path additionally contains \acp{SFF}, while a low strength path contains all types of gates and \acp{FF}. State \acp{FF} often have high (strength) \acp{FP} \cite{3513086}.

\subsubsection{\ac{SCC}} An \ac{SCC} is a set of connected nodes in a graph, such that: 1) there is a path from every node to every other node, 2) every node which satisfies property 1) is part of the set. In netlists, gates map to nodes, and wires map to edges in a graph. Thus, a \ac{SCC} is a set of strongly connected gates. \acp{SCC} can efficiently be identified with Tarjan's algorithm \cite{Tarjan1972DepthFirstSA}. We define special \acp{SCC} of a netlist: the \textit{\ac{FSM} \ac{SCC}} contains all or the majority of \acp{SFF} of the original \ac{FSM}, and the \textit{\ac{FSM-HP} \ac{SCC}} contains all or the majority of \acp{SFF} of the \ac{FSM-HP}. However, due to \ac{SCC}-property 2), both special \acp{SCC} might contain additional \acp{FF}. All other multi-\ac{FF} \acp{SCC} are defined as data \acp{SCC}.

\subsection{FSM Extraction}
\label{sec:FSMextraction}
During gate-level sequential \ac{RE}, \ac{FSM} extraction requires multiple steps: 1) identify all \acp{SFF}; 2) identify all further netlist elements which determine the \ac{FSM} states; 3) extract the \ac{STG} \cite{mcelvain2001methods, meade2016netlist}. Steps 2) and 3) can be solved by exact approaches for given \acp{SFF} and reset state. However, identifying the correct \acp{SFF} is a more challenging task, for which currently only heuristic approaches exist. Consequently, the success of retrieving a correct \ac{STG} can be measured by the success of identifying the correct \acp{SFF}.

Table \ref{tab:BackgroundOverview} provides an overview of state-of-the-art methods to identify \acp{SFF} from a reverse-engineered gate-level netlist, and
an overview of their applied features: high \ac{FP}; \ac{FP} of any strength; dependency on the same clock, reset, or enable signal; influence on the design's control signals; dissimilar gate-level structures or functions of their input cones; type or level of influence or dependency between them, ranging from loosely connected to strongly connected, e.g. \acp{SCC}; further structural features, like gate types in the input cone.
The table shows each method's features for identifying \acp{SFF}, thus showing methods with similar or frequently applied features.

To the best of our knowledge, the methods \cite{mcelvain2001methods, shi2010highly} identify \acp{SFF} to be \acp{FF} with a high \ac{FP} \cite{mcelvain2001methods} or \acp{FF} with a high \ac{FP} and an influence on control signals \cite{shi2010highly}. \Textcite{shi2010highly} additionally group \acp{FF} based on enable signals and \acp{SCC}.
The authors of RELIC \cite{meade2016gate} classify \acp{FF} into state and non-state \acp{FF} based on the dissimilarity of their input cones. \acp{FF} with less similar input cones are classified as \acp{SFF}. With grouping, authors in \cite{brunner2019improving} and \cite{meade2018old} improved the approach's performance. To improve the results, the work in \cite{brunner2019improving} also checks a potential \ac{SFF} for the existence of \acp{FP}. The work in \cite{3513086} introduces structural post-processing based on connectivity and path strength, while the method in \cite{9218616} replaces the structural with a functional similarity determination. The netlist analysis toolset NetA \cite{meade2019neta, netadoku} includes some implementations of RELIC.
One of these, the binary called \textit{relic}, extends the original method by a \ac{PCA} and structural features, resulting in a Z-Score value for each FF signal.
The higher the Z-Score value, the more likely it is a \ac{SFF}.
The authors of NetA suggest combining RELIC with \ac{SCC} identification (RELIC-Tarjan): RELIC identifies the most likely \ac{SFF}, i.e. the signal with the highest Z-Score value.
With Tarjan's algorithm, NetA determines all \acp{SCC} within the \acp{FF} of the netlist and selects the \ac{SCC} which contains this most likely \ac{SFF}.
Finally, it classifies all \acp{FF} of the selected \ac{SCC} as \acp{SFF}.
The topological analysis in \cite{fyrbiak2018difficulty} first groups \acp{FF} based on \ac{FF} types and enable, clock, and reset signals.
Next, it splits these groups of \acp{FF} based on existing \acp{SCC} and removes \acp{FF} if they do not have a high \ac{FP} or do not have enough influence on each other.
Finally, it removes \ac{SFF} groups if one of its \acp{SFF} does not have enough effect on control signals.
A recent method \cite{9643498} uses \acp{GNN} and structural features to identify state \acp{FF}.
The features include, for example, the number of gate types, inputs, or the betweenness centrality.
Finally, the method removes all \acp{FF} not part of an \ac{SCC}. 

\subsection{Exploitable \ac{FSM} Extraction Features}
\label{sec:WA}
By analyzing current \ac{SFF} identification methods, we identify frequently used features, like the high \ac{FP}, the dissimilarity, or the influence/dependency behavior.
Thus, these features form good target features when designing attractive \acp{FSM-HP}.
In addition, we identify two features that have a significant impact on the success of identification methods and can be avoided during \ac{FSM} design: high \ac{FP} and dissimilarity (highlighted in Table \ref{tab:BackgroundOverview}).
We show that these two features can be exploited to build unattractive \acp{FSM}.
We change \ac{FSM} designs such that not all of their \acp{SFF} possess all of these features without changing their original functionality.
As a result, \ac{SFF} identification methods that use these features will not correctly identify all \acp{SFF}, and thus, a correct \ac{RE} of unattractive \acp{FSM} will fail.

Also, adapting \ac{SFF} identification approaches to better identify unattractive \acp{FSM} is not assumed to be a promising solution.
If identification methods would use less restrictive features, such that they also identify \acp{SFF} of unattractive \acp{FSM}, the false positive rate, i.e. the number of \acp{FF} which are wrongly identified as \acp{SFF}, will drastically increase.

\section{Methodology}
\label{sec:methodology}
This section introduces the two parts of the new \ac{FSM} obfuscation: hardware \acp{FSM-HP} and unattractive \acp{FSM}.

\subsection{Hardware \ac{FSM} Honeypots}
\label{sec:FSMHP}
The first part of the new \ac{FSM} obfuscation methodology is hardware \acp{FSM-HP}, which pretend to be the correct \acp{FSM} but do not have an impact on the design. These \acp{FSM-HP} must be more attractive for sequential \ac{RE} methods than the original \acp{FSM} so that they are identified as best \ac{FSM} candidates. 
We assume no further design restrictions for \acp{FSM-HP}.

An \ac{FSM-HP} will be added to the original design, e.g. as a separate module.
To avoid easy detection due to its isolated, unconnected appearance, we use original design inputs, particularly the original design's reset and clock signal.
The outputs of the \ac{FSM-HP} should pretend to control the design behavior, e.g. by using techniques like dummy contacts \cite{chow2007integrated}, logic redundancy \cite{TriRedundancyLyons}, or primary module outputs. 
Additionally, the more of the typical \ac{FSM} features an \ac{FSM-HP} fulfills---like the ones in Table \ref{tab:BackgroundOverview}---the more attractive it becomes for state-of-the-art \ac{RE} methods.
Two highly relevant features are high \acp{FP} and connectivity.
Both features can be easily verified for a gate-level netlist representation of an \ac{FSM-HP} design. 
Further design options and two concrete \ac{FSM-HP} implementations are presented in sections \ref{ref:DesignOptions} and \ref{sec:Results}. 

\subsection{Unattractive \acp{FSM} }
\label{sec:UFSM}
The second part of the new \ac{FSM} obfuscation methodology is unattractive \acp{FSM}, which help to make \acp{FSM-HP} more attractive than the original \acp{FSM}.
An unattractive \ac{FSM} has a specific design that exploits, i.e. does not fulfill, a certain \ac{SFF} feature of one or more specific \ac{SFF} identification algorithms.
As a result, the algorithm and thus the \ac{FSM} extraction fail.
Different strategies exist to achieve such an \ac{FSM} design.
Either the designer knows the requirements and designs the \ac{FSM} accordingly, or an existing \ac{FSM} is redesigned without changing the original functionality.
The second strategy is preferred if possible, as it can be done independently of the \ac{FSM} design process.
We introduce two redesign methods to build an unattractive \ac{FSM} based on the identified features in section \ref{sec:WA}: dissimilarity and high \ac{FP}.

\subsubsection{Dissimilarity Approach}
\label{sec:UFSM_Sim}
\begin{figure}
     \centering
     \begin{subfigure}[b]{0.24\textwidth}
         \centering
         \includegraphics[width=0.8\textwidth]{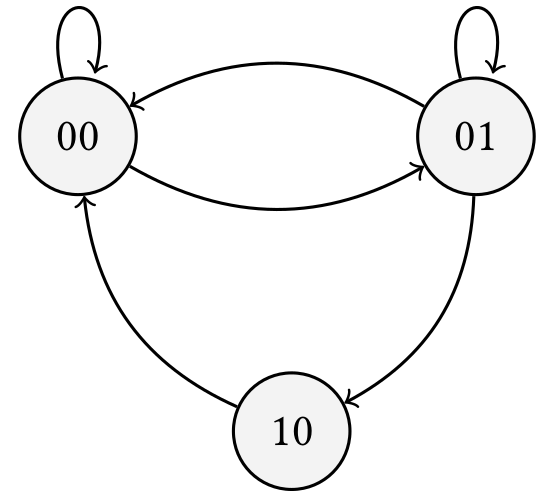}
         \caption{Original \ac{FSM}}
         \label{fig:E_Diss_orig}
     \end{subfigure}
     \hfill
     \begin{subfigure}[b]{0.24\textwidth}
         \centering
         \includegraphics[width=0.8\textwidth]{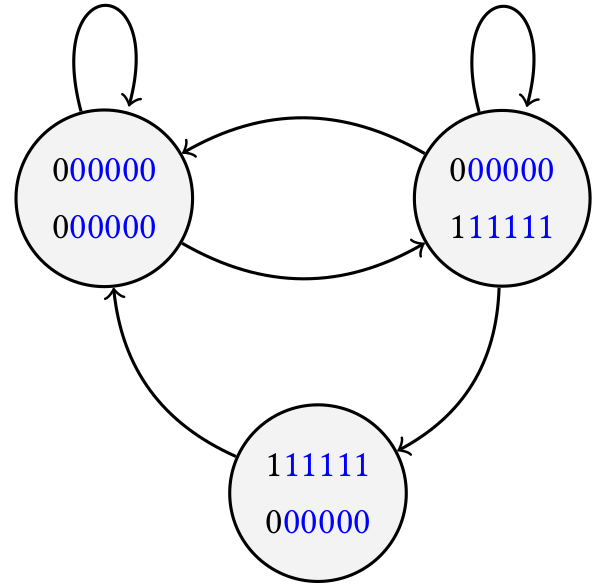}
         \caption{Unattractive \ac{FSM}}
         \label{fig:E_Diss_unattractive}
     \end{subfigure}
        \caption{State bit replication of the design fpSqrt \cite{opencoresGeneral}}
        \label{fig:E_Diss}
\end{figure}

State FFs usually have less similar input structures than data \acp{FF} because data bits are often processed similarly \cite{meade2016gate}.
Thus, an unattractive \ac{FSM} should have a low dissimilarity, i.e. a high similarity score.
One can calculate a similarity score for an \ac{FF} by comparing its \ac{FF} input structure with all other \ac{FF} input structures of the design \cite{meade2016gate}. To increase the similarity score of \ac{SFF} input structures, we replicate each state bit in the \ac{RTL} description multiple times.
The example in Fig. \ref{fig:E_Diss} shows the \ac{FSM} of the design fpSqrt (see section \ref{sec:Results_NetA}) before and after replicating each state bit five times (marked in blue).
The synthesis options are modified so that no re-encoding of the \ac{FSM} and no merging of the \acp{FF} occur, and thus the replicated state bits are translated into individual \acp{FF}.
State bit replication increases the number of required \acp{SFF}, including their input logic, but does not change the overall design functionality, as only the state
labeling is changed.
As a result, all \acp{FF} representing the replicated bits of one state bit will have increased similarity scores, making them more difficult to identify as \acp{SFF}.

In the specific case of RELIC-Tarjan \cite{meade2019neta,netadoku}, it may not be sufficient to increase the similarity of \acp{SFF} because RELIC-Tarjan uses only the signal with the highest Z-Score value to identify the corresponding set of \acp{SFF} (see section \ref{sec:FSMextraction}).
As a consequence, the identification will also succeed if any \ac{FF} of the \ac{FSM} \ac{SCC}---even if it is not a \ac{SFF}---has the highest Z-Score value.
To prevent this, one could replicate all signals from the \ac{FSM} \ac{SCC}; however, replicating state and counter bits is sufficient for most cases because they often have the highest Z-Score values.
Counter bit replication appears more challenging than state bit replication because counters usually have significantly more states than \acp{FSM}, e.g. 256 states for an 8-bit counter, and are usually not assigned within a case structure.
We developed a technique that allows a counter bit replication without using a costly case structure for the counter state assignment.
We increment or decrement an extra counter signal and then assign the replicated bits to the same value of the original counter bit.
This method is valid as long as no counter over- or underflow occurs. 
As an example, assume a 3-bit counter register \texttt{c}, which counts up using the following source code in the original design: 

\begin{minipage}{0.9\linewidth}
\ttfamily\small 
c <= c + 1;
\end{minipage}

\noindent We replicate each counter bit twice and adjust the calculation by adding a 9-bit temporary \texttt{c\_t} and following code lines:

\vspace{.5\baselineskip}
\begin{minipage}{0.9\linewidth}
\ttfamily\small
c\_t = c + 1;\\
c[8:6] <= (c\_t[8:6]==3'b001)?3'b111:c\_t[8:6];\\
c[5:3] <= (c\_t[5:3]==3'b001)?3'b111:c\_t[5:3];\\
c[2:0] <= (c\_t[2:0]==3'b001)?3'b111:c\_t[2:0];
\end{minipage}
\vspace{.5\baselineskip}

\noindent Similar to state bit replication, the design's functionality is unaffected, as only the counter state labeling changes.
The similarity score of the \acp{FF} belonging to the counter will increase, hindering the identification of the correct \ac{FSM} \ac{SCC}. 

\subsubsection{FP Approach}
\label{sec:UFSM_FP}
State FFs usually have a high \ac{FP}. This was one of the first features used for \ac{SFF} identification  \cite{mcelvain2001methods,shi2010highly} and is still used in recent approaches \cite{fyrbiak2018difficulty}.
Recently, an \ac{FSM} obfuscation method was published which requires \acp{SFF} without high \acp{FP} to apply a camouflaging technique \cite{brunner2022TimingCamouflage}.
Thus, the work in \cite{brunner2022TimingCamouflage} developed two methods to avoid a high \ac{FP} for a \ac{SFF} without changing its original functionality by redesigning an \ac{FSM}: $R_A$ which is applied on one-hot-encoded \acp{FSM} and on gate-level netlists, and $R_B$ which is applied on binary-encoded \acp{FSM} and on \ac{RTL} code.
$R_A$ partially disconnects a \ac{SFF} from posterior logic in such a way that the high \ac{FP} is removed. Where disconnected, the signal is replaced by a Boolean function that outputs one if all other \ac{SFF} values equal zero---the definition of one-hot encoding.
$R_B$ adds extra, dummy transitions to the \ac{RTL} description of the \ac{FSM} and
controls them by an obfuscation signal. The dummy transitions enable one of the \ac{FSM} bits to be determined without considering its own value, i.e. using only the other \ac{FSM} bit values and inputs.
As a result, both methods ensure that at least one of the \acp{SFF} does not influence itself within one clock cycle.
Consequently, this \ac{SFF} is free of high \acp{FP} while the original \ac{FSM} functionality does not change.
We adopt these techniques using the resulting, redesigned \ac{FSM} as unattractive \ac{FSM}.
State \ac{FF} identification methods that use high \acp{FP} as a feature will no longer identify this \ac{SFF} with medium or low \ac{FP}.
This results in an unsuccessful \ac{SFF} identification and \ac{FSM} extraction.

\subsection{Design Options and Complexities}
\label{ref:DesignOptions}
The applied redesign method sets the design options and complexities to generate unattractive \acp{FSM}.
While the dissimilarity approach is implemented by hand, the redesign method of the \ac{FP} approach can be partly automated and was shown to have short runtimes, see the results in \cite{brunner2022TimingCamouflage}. 
In contrast to unattractive \acp{FSM}, various options exist for designing an \ac{FSM-HP}.
The work aims to show this variety of design options rather than to give plenty of individual design instructions.
\ac{FSM-HP} can be built on \ac{RTL} or as gate-level netlists, by hand or automatically by a generator, with identical or changed synthesis options.
Each design option has different pros and cons. 
Designing on \ac{RTL} level or by hand allows a user-defined functionality, leading the attacker to targeted wrong conclusions about the design.
Designing on a gate-level netlist enables better control over the final netlist structure because the synthesis will not determine the gate representation itself. Better control can ease the achievement of attractive gate-based features, like dissimilarity.
An automatic generation can create a high number of \acp{FSM-HP} variations in a short time.
Separate synthesis processes allow the designer to maintain all design-specific synthesis settings for the original design while choosing suitable settings for the \ac{FSM-HP} to achieve maximum attractiveness.
The complexity of generating an \ac{FSM-HP} is independent of the remaining design.
Thus, the complexity does not change whether the \ac{FSM-HP} is generated for a toy example or an industrial design.
With automatic \ac{FSM-HP} generators, the runtime to build an \ac{FSM-HP} can be negligibly small.

\section{Results}
\label{sec:Results}
We demonstrate the different obfuscation approaches using nine open-source designs: from OpenCores \cite{opencoresGeneral} (aes\_core, altor32\_lite, fpSqrt, gcm\_aes, an adapted uart), cryptography designs \cite{secworksGeneral} (sha1\_core, siphash), a submodule and a complete core of a RISC-V processor (mem\_interface \cite{fsmMem}, picorv32 \cite{picorv32}).
For synthesis, the open-source tools qflow \cite{qflow} and yosys \cite{yosys} are applied without adding specific timing constraints.
Table \ref{tab:benchmarks} provides additional information: the number of \acp{FF}, of multi-\ac{FF} \acp{SCC}, of \acp{FSM}, of \acp{SFF} per \ac{FSM}, and the encoding of the \ac{FSM} states after synthesis.
The synthesis setup adjusts all identified \acp{FSM}, creating a one-hot encoding for most designs. Three designs, $\alpha)$, $\beta)$, and $\iota)$, consist of only a single source code module, while all others comprise a minimum of two modules. The use of the 32-bit RISC-V core demonstrates the adaption for realistic designs.
We evaluate the obfuscation results using two \ac{SFF}  identification approaches: RELIC-Tarjan \cite{netadoku} and the topological analysis \cite{fyrbiak2018difficulty}, see section \ref{sec:FSMextraction}. The \ac{SFF} identification is considered to be successful if 100\% sensitivity is achieved, i.e. if all \acp{SFF} are correctly identified, regardless of how many non-\acp{SFF} are wrongly identified.
To evaluate the obfuscation approach, we differ between the successful identification of \acp{SFF} of the unattractive \ac{FSM} and the successful identification of \acp{SFF} of the \ac{FSM-HP}.

\begin{table}[t]
    \caption{Single and multi-module designs and their number of \acp{FF}, multi-\ac{FF} \acp{SCC}, \acp{FSM}, \acp{SFF} and the type of encoding when synthesizing with default optimization settings}
    \centering
    \label{tab:benchmarks}
    \begin{tabular}{ll|ccccc}
        & Design & \#\acp{FF} & \#\acp{SCC} & \#\acp{FSM}  & \#\acp{SFF}  & encoding \\
        & & & &  & (per FSM) & \\
        \hline
        $\alpha)$ & uart & 64 & 2 & 2 & 3,2 & binary\\
        $\beta)$ & mem\_int. & 75 & 1 & 1 & 7 & one-hot \\
        \hdashline
        $\gamma)$ & siphash & 794 & 2 & 1 & 8 & one-hot \\
        $\delta)$ & sha1\_core & 850 & 3 & 1 & 3 & one-hot \\
        $\epsilon)$ & aes\_core & 901 & 2 & 1 & 16 & one-hot \\
        $\zeta)$ & altor32\_l & 1249 & 2 & 1 & 6 & one-hot \\
        $\eta)$ & fpSqrt & 1331 & 2 & 1 & 3 & one-hot\\
        $\theta)$ & gcm\_aes & 1697 & 5 & 1 & 10 & one-hot \\
        \hdashline
        $\iota)$ & picorv32 & 1598 & 1 & 2 & 4,7 & one-hot \\
    \end{tabular}
\end{table}

\begin{figure}
    \centering
	\includegraphics[width = 0.46\textwidth]{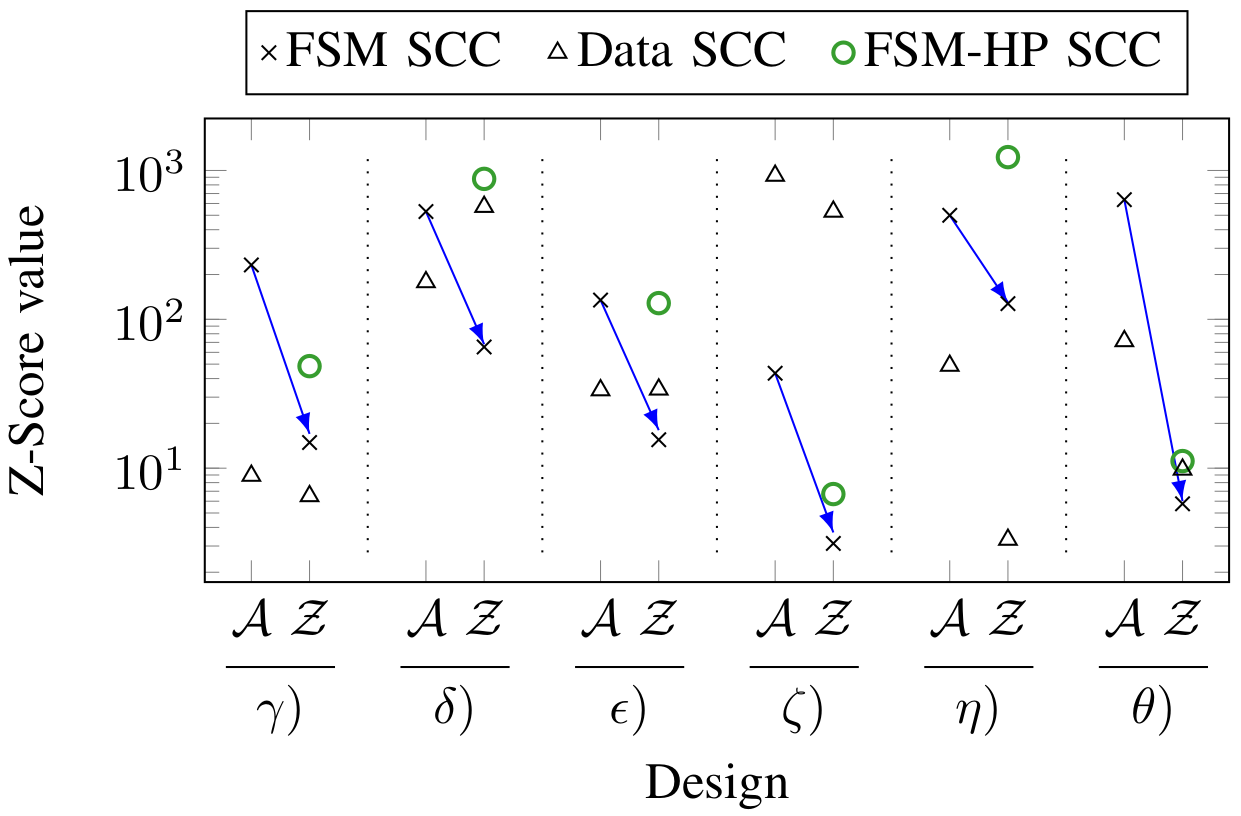}
	 \caption{Maximum Z-Score value of the \acp{FF} in any data \ac{SCC} and of the \acp{FF} in the \ac{FSM} \ac{SCC} before ($\mathcal{A}$) and after ($\mathcal{Z}$) obfuscation, and maximum Z-Score value of the \acp{FF} in the \ac{FSM-HP} \ac{SCC}}
    \label{fig:ResultNeta}
\end{figure}
\subsection{RELIC-Tarjan}
\label{sec:Results_NetA}
This section shows the successful obfuscation, using the dissimilarity approach combined with an \ac{FSM-HP}, against similarity-based \ac{SFF} identification.

We use designs with one \ac{FSM} and at least two multi-\ac{FF} \acp{SCC}, i.e. designs $\gamma) - \theta)$, see Table \ref{tab:benchmarks}.
For each design, we synthesize the original and the obfuscated version.
For comparability, we also deactivate the re-encoding when synthesizing the original netlists, see section \ref{sec:UFSM_Sim}.
As a result, some \acp{FSM} of Table \ref{tab:benchmarks} remain binary-encoded instead of being changed to one-hot encoding.
We then evaluate the obfuscation by applying the RELIC-Tarjan approach (\textit{relic} and \textit{tjscc} from NetA) on both netlists.
The tool \textit{relic} is used with its default settings and the option \textit{buf}.

The dissimilarity approach replicates state bits and, if necessary, other signals of the \ac{FSM} \ac{SCC}, three, five, or 31 times, dependent on the achieved Z-Score reduction.
To design an \ac{FSM-HP}, we copy the original \ac{FSM} source code, modify its next state or output logic, and use its outputs as primary module outputs.
This keeps realistic \ac{FSM} features.
For some designs, this process had to be repeated to receive an \ac{FSM-HP} \ac{SCC} element with a Z-Score value higher than those of the \ac{FSM} \ac{SCC}.
We recognized an increased challenge to find such a suitable \ac{FSM-HP} if the original \ac{FSM} has few state bits, e.g. $2$, or a strong cyclic behavior.
We assume that the \acp{SFF} of such \acp{FSM} have features besides the similarity score that are well identifiable by the tool \textit{relic}.
Note that it is realistic to assume that designers can evaluate their \ac{FSM-HP} or unattractive \ac{FSM} design, e.g. regarding Z-Score values: a designer also can access and apply \ac{RE} tools.

Fig. \ref{fig:ResultNeta} plots the maximum Z-Score value of the \acp{FF} in any data \ac{SCC} and the maximum Z-Score of the \acp{FF} in the \ac{FSM} \ac{SCC} using the original netlists ($\mathcal{A}$), compared against the maximum Z-Score value of the \acp{FF} in any data \ac{SCC}, the maximum Z-Score value of the \acp{FF} in the \ac{FSM} \ac{SCC}, and the maximum Z-Score of the \acp{FF} in the \ac{FSM-HP} \ac{SCC} using the obfuscated netlists ($\mathcal{Z}$).
The obfuscation succeeds for all designs, as an \ac{FSM-HP} could always be designed such that at least one \ac{FF} in its \ac{SCC} has a higher Z-Score value than any \ac{FF} of the \ac{FSM} \ac{SCC} (compare the circles and the crosses for the obfuscated netlists $\mathcal{Z}$).
Due to the dissimilarity approach, for all designs, the maximum Z-Score value of the \acp{FF} in an \ac{FSM} \ac{SCC} decreased for the obfuscated design.
We highlight this change with blue arrows.
For the majority of designs, we achieve a maximum Z-Score value for the \acp{FF} of the original \ac{FSM} \ac{SCC} that is below the maximum Z-Scores for the \acp{FF} of data \acp{SCC} and the \ac{FSM-HP} \ac{SCC}.
Note that Fig. \ref{fig:ResultNeta} does not show all, but only the maximum Z-Score values.
Otherwise, there would be further plot points below the cross, the triangle, or the circle.
The ranking position of the maximum Z-Score value of the \ac{FSM} \ac{SCC} should not be predictable so as not to open up new attack possibilities.
Adding more than one \ac{FSM-HP} to the design supports a not predictable ranking and, thus, strengthens the obfuscation.
Summarizing, the RELIC-Tarjan procedure will favor the \ac{FSM-HP} over the unattractive \ac{FSM}, allowing a successful \ac{FSM} obfuscation.

\begin{table}[t]
    \let\TPToverlap=\TPTrlap
    \caption{Topological analysis with and without obfuscation (\checkmark: all correct \acp{SFF}  identified, ($x$/$y$): $x$ out of $y$ \acp{SFF}  correctly identified resulting in a successful obfuscation)}
    \centering
    \label{tab:TopAn}
\begin{threeparttable}
    \begin{tabular}{l|c:ccc}
        Design & without obfuscation& \multicolumn{2}{c}{with \ac{FP} approach and \ac{FSM-HP}}\\
         & orig. \ac{FSM}  & orig. \ac{FSM}  & \ac{FSM-HP}  \\
        \hline
        $\alpha)$ & \checkmark & (1/2), (2/3) & \checkmark \\
        $\beta)$ & \checkmark & (0/7) & \checkmark \\
        $\gamma)$ & \checkmark & (7/8) &  \checkmark\\
        $\iota)$ \tnote{1} & \checkmark & (3/4) &  \checkmark\\
    \end{tabular}
\begin{tablenotes}[flushleft]\footnotesize
       \item [1] \ac{FSM} of the memory interface
    \end{tablenotes}
\end{threeparttable}
\end{table}
\subsection{Topological Analysis}
\label{sec:Result_top}
This section shows the successful obfuscation using the \ac{FP} approach combined with an \ac{FSM-HP} against topological-analysis-based \ac{SFF} identification. We use four designs of Table \ref{tab:benchmarks} for which the topological analysis can extract an \ac{FSM} candidate with all original \acp{SFF}, see column 2 of Table \ref{tab:TopAn}.
We slightly adapt the final step in our implementation of the topological analysis because otherwise, our obfuscation worked too easily:
Instead of removing the \ac{FSM} candidate as a whole if one \ac{FF} does not show any control behavior \cite{fyrbiak2018difficulty}, we only remove the \ac{FF} itself.
\footnote{Due to performance limitations of our topological analysis implementation, for the picorv32, the output control behavior calculation (last step) could not be finished.
Thus, Table \ref{tab:TopAn} shows the results for picorv32 without performing this last step.
However, the actual obfuscation results are assumed to improve further because this last step would remove additional \acp{FF} of \ac{FSM} candidates.}

For the \ac{FP} approach, we apply an \ac{FSM} redesign method on an arbitrary state bit or \ac{SFF} of the original \ac{FSM}: $R_A$ for one-hot-encoded, and $R_B$ for binary-encoded designs.

When applying the \ac{FSM-HP}, in contrast to the demonstration in section \ref{sec:Results_NetA}, the same \ac{FSM-HP} is added to all designs; only the inputs are changed to match the inputs of the original design.
The \ac{FSM-HP} has five state bits and fulfills the features of the topological analysis as best as possible, including \acp{FP}, an \ac{SCC}, and good influence/dependency and control behavior. 
Finally, topological analysis identifies no or only some \acp{SFF} of the original \ac{FSM}, but all \acp{SFF} of the \ac{FSM-HP}, see columns 3 and 4 of Table \ref{tab:TopAn}.
This leads to the following two possible results:
\begin{itemize}
    \item No \ac{FSM} candidate contains any \acp{SFF} of the original design. As a result, data \acp{FF} or the \ac{FSM-HP} \acp{FF} will be used to extract an incorrect \ac{FSM}.
    \item An \ac{FSM}  candidate contains parts of the original \acp{SFF}. As a result, these \acp{SFF} or data \acp{FF} or \ac{FSM-HP} \acp{FF} will be used to extract an incorrect \ac{FSM}.
\end{itemize}
Both cases successfully prevent the correct extraction of the original \ac{FSM}.

\subsection{Overhead}
\label{sec:Results_overhead}
On average, the obfuscated designs in section \ref{sec:Results_NetA} have 51\%, the obfuscated designs in section \ref{sec:Result_top} have 8\% more cell area.
Thus, the average overheads are larger and smaller than the ones in \cite{rahman2022practical} (24\%), or both smaller than the ones in \cite{keyless2023} (288\%).
We assume that the large overhead of RELIC-Tarjan comes from the replicated logic of the dissimilarity approach and the applied \ac{FSM-HP} insertion technique, i.e. copying and modifying the original \ac{FSM}.
However, as the obfuscation targets the \ac{FSM} and an \ac{FSM} is usually the smallest part of a design, our measured overhead results should decrease for larger industrial designs.
In contrast to the area overhead and in contrast to recent obfuscation methods \cite{rahman2022practical, keyless2023}, for our designs, on average, the slack time is not affected: -0.7\% for the obfuscated designs in section \ref{sec:Results_NetA} and -2\% for the obfuscated designs in section \ref{sec:Result_top}.
We used proprietary EDA tools to perform the measurements, assuming a frequency of \SI{20}{\mega\hertz}.

\section{Conclusion}
\label{sec:Conclusion}
The work presents a two-part \ac{FSM} obfuscation approach to prevent sequential \ac{RE}: hardware \acs{FSM} honeypots and unattractive \acp{FSM}.
Using one similarity-based and one topological-analysis-based \ac{SFF} identification method, we demonstrate that state-of-the-art \ac{RE} tools favor the more attractive \acp{FSM-HP} or cannot correctly identify the unattractive original \acp{FSM}.
This successfully obfuscates the original \acp{FSM} and allows control over what will be identified by the attacker.

The novel obfuscation approach is extendable by investigating other \ac{RE} tool features and exploiting them for unattractive \acp{FSM}.
Also, the obfuscation can be strengthened if more than one \ac{FSM-HP} is added to a design.
In addition, the generation of \acp{FSM-HP} and unattractive \acp{FSM} can be adapted to new identification mechanisms with new features.
Thus, the obfuscation approach has the potential to be secure also for novel identification techniques.

\printbibliography

\end{document}